\documentclass[11pt]{article}  
 \newcommand{\bc}{\begin{center}}
 \newcommand{\ec}{\end{center}}
                   \newcommand{\bfr}{\begin{flushright}}
                   \newcommand{\efr}{\end{flushright}}
   
     \newcommand{\be}{\begin{enumerate}}
     \newcommand{\ee}{\end{enumerate}}
        \newcommand{\bi}{\begin{itemize}}
        \newcommand{\ei}{\end{itemize}}
            \newcommand{\bd}{\begin{description}}
            \newcommand{\ed}{\end{description}}
                \newcommand{\beq}{\begin{equation}}
                \newcommand{\eeq}{\end{equation}}
                  \newcommand{\bea}{\begin{eqnarray}}
                  \newcommand{\eea}{\end{eqnarray}}

      \newcommand{\bfi}{\begin{figure}}
      \newcommand{\efi}{\end{figure}}
\newcommand{\bay}{\begin{array}{l}}
\newcommand{\eay}{\end{array}}
            \newcommand{\dd}{\mbox{d}}

    \newcommand{\pa}{\partial}
    \newcommand{\del}{\delta}
    \newcommand{\Del}{\Delta}
    
    \newcommand{\la}{\lambda}
    
    \newcommand{\al}{\alpha}
    \newcommand{\sig}{\sigma}
    \newcommand{\eps}{\epsilon}
    
    \newcommand{\Ga}{\Gamma}
    
    \newcommand{\Om}{\Omega}

\newcommand{\xx}{\mbox{\boldmath $x$}}

 \usepackage{amsmath}
  \usepackage{amsthm}

\usepackage{enumitem}
\usepackage{graphicx}
\usepackage{comment}
\usepackage{cases}
\usepackage{bm}
\usepackage[dvipsnames]{xcolor}
\usepackage{algorithm} 
\usepackage{algpseudocode}

\usepackage[T1]{fontenc}
\usepackage[utf8]{inputenc}
\usepackage[a4paper, textwidth=18cm, textheight=25cm]{geometry}

\usepackage{amsmath}
\usepackage{mathtools}
\usepackage{graphicx}
\usepackage{bm}
\usepackage{xcolor}
\usepackage{lineno}
\usepackage[unicode]{hyperref}
\usepackage[colorinlistoftodos]{todonotes}
\usepackage{upgreek}
\usepackage{sectsty}
\usepackage{blkarray}
\usepackage{amsfonts}

\usepackage{biblatex}
\newcommand{\footremember}[2]{%
	\footnote{#2}
	\newcounter{#1}
	\setcounter{#1}{\value{footnote}}%
}
\newcommand{\footrecall}[1]{%
	\footnotemark[\value{#1}]%
}

\begin{document}

\title{Application of $J$-Integral to a~Random Elastic Medium}
\author{Jan Eliáš\footremember{Brno}{Brno University of Technology, Faculty of Civil Engineering, Brno, Czechia} \and Josef Martinásek\footrecall{Brno} \and Jia-Liang Le\footremember{UMN}{Department of Civil, Environmental, and Geo- Engineering, University of Minnesota, MN, USA}\footremember{email}{Corresponding author: jle@umn.edu}}
\date{}

\maketitle

\section*{Abstract} This study investigates the use of the $J$-integral to compute the statistics of the energy release rate of a~random elastic medium. The spatial variability of the elastic modulus is modeled as a~homogeneous lognormal random field. Within the framework of Monte Carlo simulation, a~modified contour integral is applied to evaluate the first and second statistical moments of the energy release rate. These results are compared with the energy release rate calculated from the potential energy function. The comparison shows that, if the random field of elastic modulus is homogenous in space, the path independence of the classical $J$-integral remains valid for calculating the mean energy release rate.  However, this path independence does not extend to the higher order statistical moments. The simulation further reveals the effect of the correlation length of the spatially varying elastic modulus on the energy release rate of the specimen.

\section{Introduction}
Over the past century, linear elastic fracture mechanics (LEFM) has been developed into a~complete subject that plays a~pivotal role in a~variety of engineering disciplines. The cornerstone concept in LEFM is the energy release rate proposed by Griffith \cite{Gri21}, which provided a~definitive answer to the fundamental question on why structures with a~preexisting crack would have a~finite load carrying capacity even though the elastic stress at the crack tip is infinite. Within the framework of LEFM, various methods have been developed to calculate the energy release rate. Among them, the most prominent one is Rice's $J$-integral \cite{Ric68}. The $J$-integral has a~deep mathematical root, which can be traced back to the celebrated Noether's work on the invariant variational principle \cite{Noe1918}. Meanwhile, it has been well understood that this integral is closely related to the Eshelby tensor \cite{Esh51,Esh75,HonHer97,BalRoy16}, a~fundamental concept leading to the derivation of configurational forces for a~broad class of problems in solid mechanics \cite{Gur08}.  It is also noted that the $J$-integral can be applied to nonlinear elasticity, which provides a~good approximation to the yielding zone of plastic-hardening metals as long as there is no unloading \cite{And03,BazLe-21}. 

The most salient property of the $J$-integral is its path independence. This property allows one to choose any convenient path to calculate the energy release rate. However,  the path independence property is valid only if the material is homogeneous at least in the direction of the crack \cite{Ric68}. It has been shown that for non-homogeneous materials the original form of the $J$-integral needs to be modified in order to preserve the path independence property \cite{Eis87,HonHer97,KimPau03}. The modified $J$-integral has successfully been used to calculate the stress intensity factors for various non-homogeneous materials, such as the functionally graded materials \cite{KimPau03}. 

Many engineering structures are composed of  statistically homogeneous materials, in which the elastic properties could exhibit some degree of spatial randomness and at any point they share the same probability distribution. Some typical examples include concrete, rock, particulate composites, and ceramics.  Due to the intrinsic randomness of material properties,  the energy release rate of a specimen would be random. Quantifying the statistics of the energy release rate is an~essential step for the estimation of the randomness of the fracture energy, which is a~key material property for failure analysis. However, so far few studies have applied the $J$-integral to statistically homogeneous materials. 

Based on the theory of random fields, the spatial randomness of the elastic properties of statistically homogeneous materials can be mathematically represented by homogeneous random fields, and the statistics of the energy release rate can then be calculated through the Monte Carlo simulations. It is evident that, for each realization of the random fields of material properties, the classical $J$-integral is inapplicable since the material is inhomogeneous. On the other hand, it is noted that,  in an~average sense,  the material is homogeneous. It is reasonable to expect that Rice's $J$-integral might still be applicable for calculating the mean value of the energy release rate. However, this conjecture has not been proven mathematically. Furthermore, the consideration of spatial randomness of elastic modulus would naturally introduce the correlation length to the problem. The effect of this length scale on the overall statistics of the energy release rate including its mean value is largely not understood. To answer these questions, this study conducts  theoretical and numerical investigations on the application of  $J$-integral to a~random elastic media, which is highly relevant to the analysis of fracture behavior of large-scale structures made of brittle heterogeneous materials.  

\section{Theoretical Consideration}
Consider two-dimensional analysis of a~point-wise isotropic elastic medium with a~preexisting crack. The elastic stiffness tensor at any local point  is expressed by $C_{ijkl} (x_1, x_2)= \mu(x_1, x_2) (\del_{ik}\del_{jl} + \del_{il} \del_{jk}) + \la (x_1, x_2) \del_{ij} \del_{kl}$, where $x_1, x_2 =$ local coordinate, $\mu, \la =$ spatially varying elastic constants, $\del_{mn} =$  Kronecker delta. The elastic constants $\mu$ and $\la$ are related to the Young modulus $E$ and Poisson ratio $\nu$ by $\mu = E/2(1+\nu)$ and $\la = E\nu/(1+\nu)(1-2\nu)$. In this study, the spatial randomness of the elastic stiffness tensor is governed by a~homogeneous random field of Young modulus $E(x_1, x_2)$. The Poisson ratio is considered to be deterministic. 

\begin{figure}[!tb]
\centering\includegraphics[width=7cm]{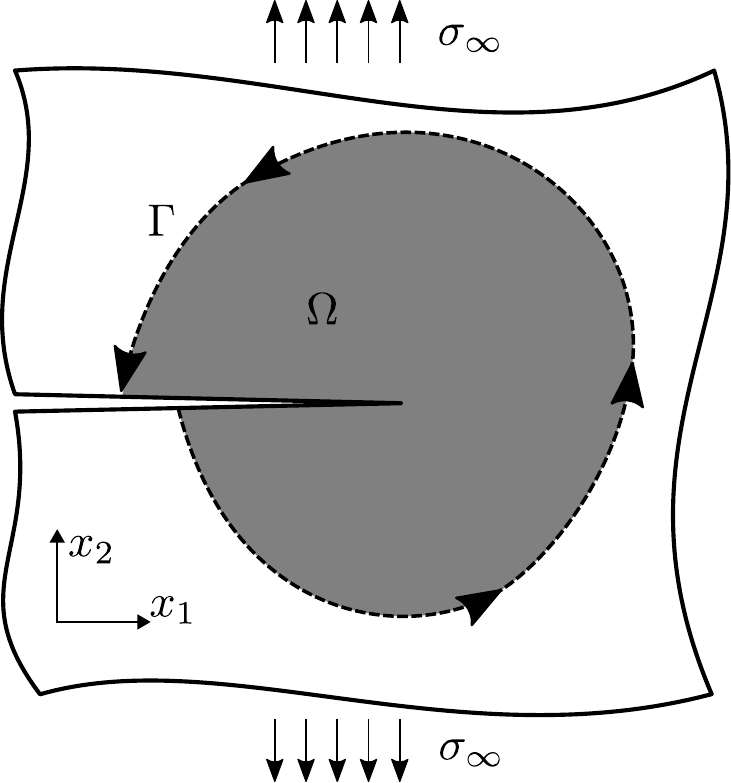}
\caption{Calculation of the energy release rate from an~arbitrary domain containing the crack tip.} \label{fig:Jbasic}
\end{figure}

Following \cite{Ric68b}, let $\Om$ be a~finite region enclosing the crack tip, and the boundary of region $\Om$ is denoted by $\Ga$ (Fig. \ref{fig:Jbasic}). The energy release rate can be written by 
\beq\label{e1}
{\cal{G}} = \lim_{\Del a \rightarrow 0} \frac{1}{\Del a} \left\{ \int_\Ga  (t_i^0 + \Del t_i) \Del u_i \dd s- \int_\Om  \left[\bar{U}(\eps_{mn}^0 + \Del \eps_{mn}) - \bar{U}(\eps_{mn}^0) \right] \dd x_1 \dd x_2 \right\}
\eeq
where $t_i^0 =$ original traction on the boundary $\Ga$ at crack length $a_0$, $\Del t_i =$  traction increment as the crack extends by an~amount of $\Del a$, $\Del u_i =$ displacement increment on $\Ga$, $\eps_{mn}^0 =$ original strain field, $\Del \eps_{mn} =$ strain increment, and $\bar{U} =\frac{1}{2} C_{ijkl} \eps_{ij} \eps_{kl} =$ strain energy density. 

To evaluate  Eq.~\eqref{e1}, we introduce a~moving coordinate system $(X_1, x_2)$,  in which the origin is attached to the crack tip. Therefore, we have $X_1 = x_1 - a$, where $ a=$ crack length, and the total differential of a~physical quantity with respect to crack length $a$ is given by $\dd (\cdot) /\dd a = \pa (\cdot)/\pa a - \pa(\cdot) /\pa X_1 =\pa (\cdot)/\pa a - \pa(\cdot) /\pa x_1$. The line integral in Eq.~\eqref{e1} can then be written by
\beq\label{e2}
\lim_{\Del a \rightarrow 0} \frac{1}{\Del a} \int_\Ga  (t_i^0 + \Del t_i) \Del u_i \dd s = \int_\Ga t_i^0(X_1, x_2)|_{a_0}  \left(\frac{\pa u(X_1, x_2)}{\pa a}\big|_{a_0} - \frac{\pa u(X_1, x_2)}{\pa X_1}\big|_{a_0} \right) \dd s
\eeq

For the domain integral in Eq.~\eqref{e1}, we first note that the difference in strain energy density is strictly due to the change in strain as the crack extends. However, when we perform the total differentiation of the strain energy density function, the term $\pa \bar{U}/\pa x_1$ includes the change in strain energy density due to the spatial variation of the elastic constants. Therefore, we need to subtract this contribution from the total differentiation, which leads to
\beq\label{e3}
\lim_{\Del a \rightarrow 0} \frac{1}{\Del a} \int_\Om \left[\bar{U}(\eps_{mn}^0 + \Del \eps_{mn}) - \bar{U}(\eps_{mn}^0) \right] \dd x_1 \dd x_2  = \int_\Om \left\{\frac{\pa \bar{U}(X_1, x_2)}{\pa a}\big|_{a_0} - \left[\frac{\pa \bar{U}(X_1, x_2)}{\pa X_1}\big|_{a_0} - \frac{1}{2} C_{ijkl, 1} \eps_{ij}\eps_{kl}|_{a_0} \right] \right\} \dd X_1 \dd x_2
\eeq
where $C_{ijkl,1} =  \pa C_{ijkl}/\pa x_1= \pa C_{ijkl}/\pa X_1$. 

Note that $\pa \bar{U}/\pa a = \sig_{ij} \pa \eps_{ij}/\pa a$. With the strain-displacement relation $\eps_{ij} = \frac{1}{2} (u_{i, j}+u_{j,i})$ and the equilibrium equation $\sig_{ij, j} = 0$, it is easy to show that $\pa \bar{U}/\pa a = (\sig_{ij} \pa u_i/\pa a),_{j}$. By applying the divergence theorem, we have
\beq\label{e4}
 \int_\Om \frac{\pa \bar{U}(X_1, x_2)}{\pa a}\big|_{a_0} \dd X_1 \dd x_2 = \int_\Ga \sig_{ij}(X_1, x_2)n_j  \frac{\pa u(X_1, x_2)}{\pa a} \big|_{a_0} \dd s =  \int_\Ga t_i^0(X_1, x_2)  \frac{\pa u(X_1, x_2)}{\pa a} \big|_{a_0} \dd s 
\eeq
where $n_j =$ component of the unit vector of the outward normal of the contour.  By substituting Eqs. \eqref{e2} and \eqref{e3} into Eq.~\eqref{e1} and considering Eq.~\eqref{e4}, we reach the final expression of the energy release rate:
\beq\label{e5}
{\cal{G}} =\int_\Ga \bar{U} \dd x_2 - \int_\Ga t_i u_{i,1} \dd s - \frac{1}{2} \int_\Om  C_{ijkl, 1} \eps_{ij}\eps_{kl}  \dd x_1 \dd x_2
\eeq
Since all the quantities in Eq.~\eqref{e5} refer to the original state, we drop the superscript ``0'' and switch back to the fixed coordinate system $(x_1, x_2)$. Eq.~\eqref{e5} is a~general expression of the energy release rate for linear elastic non-homogeneous materials using the contour integral approach \cite{Eis87,KimPau03}.  

\begin{figure}[!tb]
\centering\includegraphics[width=7cm]{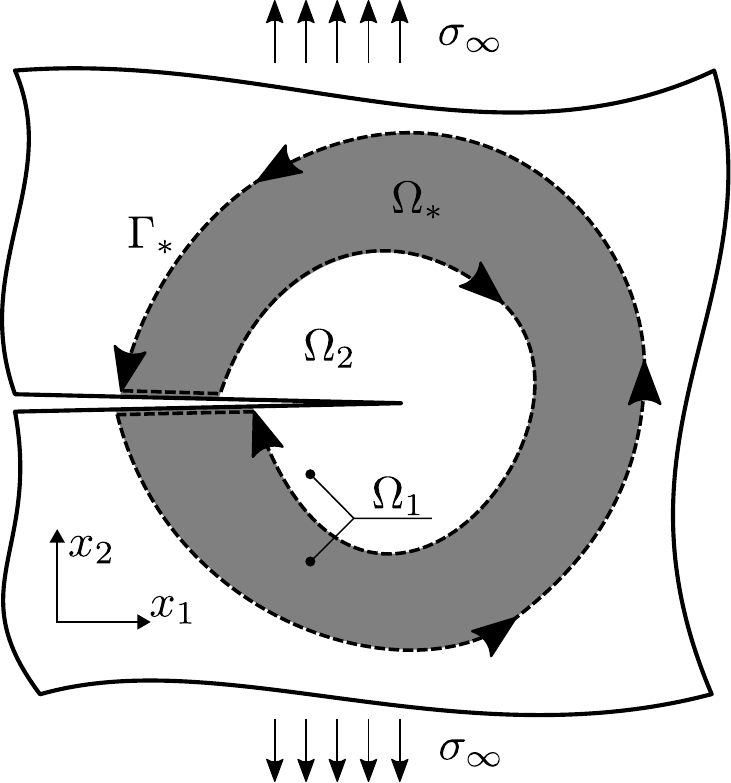}
\caption{Demonstration of the path independence property of $J_*$.} \label{fig:Jindep}
\end{figure}

Let $J_*$ denote the expression of the right hand side of Eq.~\eqref{e5}. To evaluate the dependence of $J_*$ on the choice of the domain, we consider two arbitrary choices of domains $\Om_1,~ \Om_2$ shown in Fig. \ref{fig:Jindep}. The difference between the values of $J_*$ for these two domains is given by 
\beq\label{e6}
J_*(\Om_1) - J_*(\Om_2) = \int_{\Ga^*} \bar{U} n_1 -t_i u_{i,1} \dd s - \frac{1}{2} \int_{\Om^*}  C_{ijkl, 1} \eps_{ij}\eps_{kl}  \dd x_1 \dd x_2
\eeq
where $\Om^*$ denotes the portion of domain $\Om_1$ that is outside domain $\Om_2$, and $\Ga^*$ is the boundary of $\Om^*$ including the two segments along the crack surface. By apply the divergence theorem, we can rewrite the line integral by 
\beq\label{e7}
\int_{\Ga^*} \bar{U} n_1 -t_i u_{i,1} \dd s  = \int_{\Om^*}  \bar{U}_{,1} - (\sig_{ij} u_{i,1})_{,j} \dd x_1 \dd x_2
\eeq
Note that $\bar{U}_{,1} = \frac{1}{2} C_{ijkl,1} \eps_{ij} \eps_{kl} + \sig_{ij} \eps_{ij, 1}$ and $ (\sig_{ij} u_{i,1})_{,j} = \sig_{ij} u_{i, 1j} = \sig_{ij} \eps_{ij, 1}$.  By substituting these expressions into Eq.~\eqref{e6}, we have $J_*(\Om_1) = J_*(\Om_2)$. This shows that $J_*$ is independent of the choice of the domain.

Meanwhile, we note that $J_*$ is composed of Rice's $J-$integral ($J =  \int_\Ga \bar{U} \dd x_2 - \int_\Ga t_i u_{i,1} \dd s$) and a~domain integral, i.e., $J_* = J  - \frac{1}{2} \int_\Om  C_{ijkl, 1} \eps_{ij}\eps_{kl}  \dd x_1 \dd x_2$. For a smooth field of spatially varying elastic modulus (i.e. $C_{ijkl, 1}$ is finite), the domain integral in Eq.~\eqref{e5} would approach zero as the domain shrinks to zero. Therefore, Rice's $J-$integral would yield the energy release rate if one chooses a~path that is infinitely close to the crack tip.

In the present stochastic analysis, due to the spatial randomness of elastic modulus, the value of $J_*$ for any given domain (consequently the energy release rate) is intrinsically random. The primary focus of reliability-based analysis is on the first and second moments of statistics. The mean energy release rate can be expressed by
 \beq\label{e8}
\langle J_* \rangle = \langle J \rangle -  \frac{1}{2} \int_\Om  \langle C_{ijkl, 1} \eps_{ij}\eps_{kl}\rangle   \dd x_1 \dd x_2 
\eeq
where $\langle x \rangle =$ mean value of $x$. To evaluate Eq.~\eqref{e8}, we note that, for each realization of the random field of Young modulus, the strain at a~given point depends on the loading condition, structure geometry, and spatial distribution of Young modulus. It has been shown that the homogeneous random field and its spatial gradient are uncorrelated \cite{Van10}.  Since the strain field only depends on $E(x_1, x_2)$, it would be uncorrelated to the gradient of the random field $E(x_1, x_2)$. Therefore, we have
\beq\label{e10}
  \langle C_{ijkl, 1} \eps_{ij}\eps_{kl}\rangle    =   \langle C_{ijkl, 1}\rangle \langle \eps_{ij}\eps_{kl}\rangle  
  \eeq
It is evident that $C_{ijkl, 1} \propto \pa E(x_1, x_2)/\pa x_1$. Since $E(x_1, x_2)$ is a~homogeneous random field, the mean value of $E(x_1, x_2)$ is a~constant. Therefore, the mean value of $\pa E(x_1, x_2)/\pa x_1$ must be zero.  What it follows is that $\langle C_{ijkl, 1}\rangle = 0$, and with Eq.~\eqref{e8} we have 
\beq\label{e11}
\langle J_* \rangle = \langle J \rangle 
\eeq
Based on Eq.~\eqref{e11}, we conclude that, if the spatial randomness of elastic constants can reasonably be described by a~homogeneous random field, the mean energy release rate would exactly be equal to the mean value of Rice's $J-$integral, which is independent of the choice of the contour. 

Since $J = J_*+\frac{1}{2} \int_\Om  C_{ijkl, 1} \eps_{ij}\eps_{kl}  \dd x_1 \dd x_2$ and $J_*$ is the energy release rate,  in general the second statistical moment of $J$ is path dependent. Clearly the variance of $J$ would approach the variance of the energy release rate under either of the two scenarios: 1) the domain $\Om$ is very small, i.e., the path is infinitely close to the crack tip, and 2) the autocorrelation length of the random field $E(x, y)$ is considerably larger than the size of domain $\Om$.

\section{Numerical Study}

\begin{figure}[!tb]
\centering\includegraphics[width=11cm]{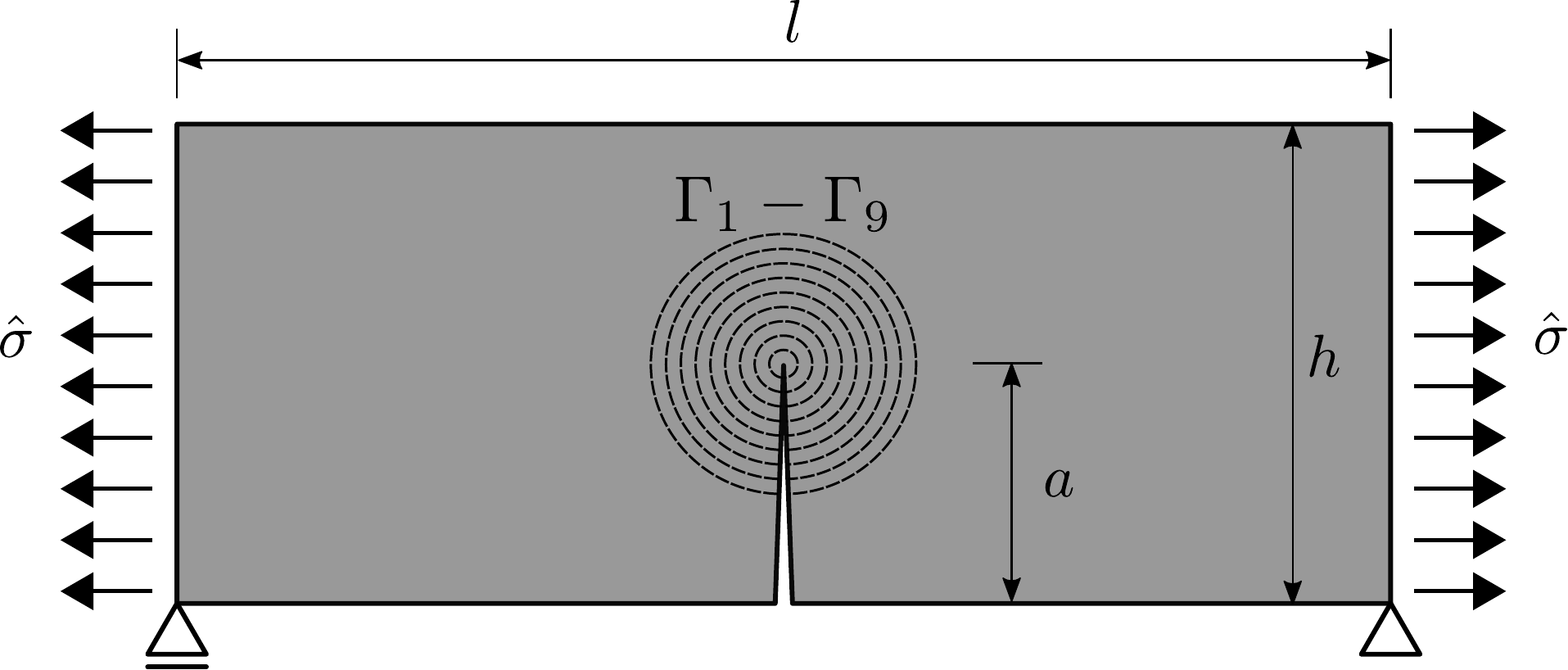}
\caption{Geometry of the SENT specimen and the different contours used for evaluation of $J$ and $J_*$.} \label{fig:geometry}
\end{figure}

To validate the foregoing analytical model, we perform a~numerical analysis on the energy release rate of a~single edge notched tensile (SENT) specimen shown in Fig.~\ref{fig:geometry}. The analysis considers a~linearly elastic material under a plane stress condition.  The specimen is loaded by a~far field stress $\hat{\sig} =1$\,MPa at both ends and the rigid body motion is prevented by constraining the two corner nodes. The specimen has a~length $l = 0.5$\,m and width $h= 0.2$\,m. The notch length is half of the width. The commercial finite element software Ansys is used to perform the stress analysis, in which the bi-quadratic plane elements are used. The energy release rate can be calculated from the complementary potential energy function, i.e.,
\beq\label{e12}
\mathcal{G}(a) = \frac{1}{b}\frac{\pa {U^*}(a)}{\pa a} \Bigr|_{\hat{\sigma}}\approx \frac{U^*(a+\del a)-U^*(a)}{b\del a} 
\eeq
where $U^* =$ complementary potential energy, $b =$ width of the specimen in the transverse direction (here we take $b=1$), $a =$ crack length, and $\del a =$ small increment of the crack length (we consider $\del a =10$\,$\mu$m  in the present analysis). Meanwhile, Rice's $J-$integral can be readily evaluated by using the built-in function of Ansys.

We first perform a~deterministic analysis to verify the numerical model, where we take Young modulus $E =30$\,GPa and Poisson ratio $\nu = 0.2$. The numerical analysis using the complementary potential energy function  (Eq.~\eqref{e12}) yields $\mathcal{G} = 83.57$\,J/m$^2$.  Meanwhile, we  calculate the $J-$integral for nine different contours shown in Fig. \ref{fig:geometry}.  The calculation shows that the average value of $J-$integral is 83.55\,J/m$^2$ with a negligibly small standard deviation (0.022\,J/m$^2$), which demonstrates the path independence property of the $J-$integral. For this deterministic analysis, the $J-$integral yields exactly the energy release rate. To further validate the present numerical analysis, we calculate the energy release rate using the Irwin relationship, i.e. $\mathcal{G} = K_I^2/E$, where $K_I =$ stress intensity factor. According to \cite{TadPar-00}, the stress intensity factor of the SENT specimen can be approximated by
\beq\label{e13}
K_I  =\hat{\sigma}  \sqrt{2h\tan\left(\frac{\pi \alpha}{2}\right)} \left\{\frac{0.752+2.02\alpha+0.37\left[1-\sin\left(\pi \alpha/2\right)\right]^3}{\cos\left(\pi \alpha/2\right)} \right\}
\eeq
where $\al = a/h =$ relative crack length. For the SENT specimen shown in Fig.~\ref{fig:geometry}, the energy release rate calculated from the Irwin relationship is 83.6\,J/m$^2$, which agrees very well with the results obtained by Eq.~\eqref{e12} and the $J-$integral.

For the stochastic analysis, the spatial randomness of the Young modulus is represented by a homogeneous  random field $\xi(\xx)$:
\begin{align}\label{e14}
E = \bar E \xi(\xx)
\end{align}
where $\bar E$ is the mean value of the modulus equal to 30\,GPa. At any given location, $\xi$ follows a~lognormal distribution function $F(\xi)$ with a~mean value of one and a~standard deviation of 0.2. The spatial correlation of $\xi$ is described by a~square exponential autocorrelation function:
\begin{align}\label{e15}
\rho(\xx_p,\xx_q) = \exp\left(-\frac{||\xx_p-\xx_q||^2}{\ell^2}\right) 
\end{align}
where $\ell =$ correlation length. In this study, different values of $\ell$ (0.02, 0.04, 0.08, and 0.16\,m) are considered to investigate the influence of the spatial correlation on the statistics of the energy release rate. 

To generate the target lognormal random field $\xi(\xx)$, we transform the random field to an~equivalent standard Gaussian random field $\eta(\xx)$. The marginal probability distribution functions of $\xi$ and $\eta$ are related by 
\begin{align}\label{e16}
\xi = F^{-1}(\Phi(\eta)) 
\end{align}
where $\Phi(y) =$ cumulative distribution function (cdf) of the  standard Gaussian variable. The isoprobabilistic transformation (Eq.~\eqref{e16}) distorts the autocorrelation structure of the random field. Therefore, we need to determine the correlation function $\hat\rho(\xx_p,\xx_q)$ of the Gaussian field $\eta(\xx)$ that would correspond to the correlation structure of the target lognormal field $\xi(\xx)$.  The correlation function $\hat\rho(\xx_p,\xx_q)$ of the equivalent Gaussian field can be obtained by the Nataf transformation of the correlation coefficient \cite{LiLu-08}.

Once the correlation function $\hat\rho(\xx_p,\xx_q)$ is obtained, we can generate the standard Gaussian field $\eta(\xx)$ by the Karhunen--Lo\`{e}ve (K-L) expansion~\cite{Kar47,Spanos1989,Ghanem2003,Ste09}:
\beq
\eta (\xx) = \sum_{k=1}^{\infty} \sqrt{\lambda_k} \zeta_k \phi_k(\xx) \label{eq:KarhunenLoeve}
\eeq
where $\zeta_k$ are independent standard Gaussian variables generated using Latin Hypercube Sampling \cite{McKaBec-79}, and $\lambda_k$ and $\phi_k(\xx)$ are the eigenvalues and eigenfunctions of the auto-covariance function given by the Fredholm integral equation of the second kind
\bea
\int_{\Omega}  \tilde\rho(\xx_p, \xx_q) \phi_k(\xx_q) \dd {\xx_q} =\lambda_k \phi_k(\xx_p)
\label{eq:fredholm}\\
\mbox{with}~~~ \int_\Om \phi_i(\xx) \phi_j(\xx) \dd \xx = \del_{ij}
\eea
Eq.~\eqref{eq:fredholm} is solved numerically. In practice, it suffice to compute only $K$ eigenmodes corresponding to the largest eigenvalues of interest, with the value of $\sum_{k=1}^K \lambda_k$ converging with a~relative error less than 1\%. 

The Gaussian random field, $\eta(\xx)$, is generated on a~sufficiently dense grid, which is independent from the finite element mesh. The Expansion Optimal Linear Estimation (EOLE) is employed to project the random field onto the integration point $\xx_g$ of the finite mesh \cite{LiKiu93}: 
\begin{align}
\eta (\xx_g) = \sum_{k=1}^{K} \frac{\zeta_k}{\sqrt{\lambda_k}}\bm{\upphi}_k^T \tilde{\mathbf{c}}_{xg} \label{eq:EOLE}
\end{align}
where $\tilde{\mathbf{c}}_{xg}$ is a~vector containing the covariances of $\eta(\xx)$ between the point $\xx_g$ and the grid points, and $\bm {\upphi}_k$ are the eigenfunctions of the covariance matrix. The EOLE method is further used to determine the $\eta$-values at points in the vicinity of the integration points for the subsequent calculation of the spatial gradient of the Young modulus.

\begin{figure}[!tb]
\includegraphics[width=\textwidth]{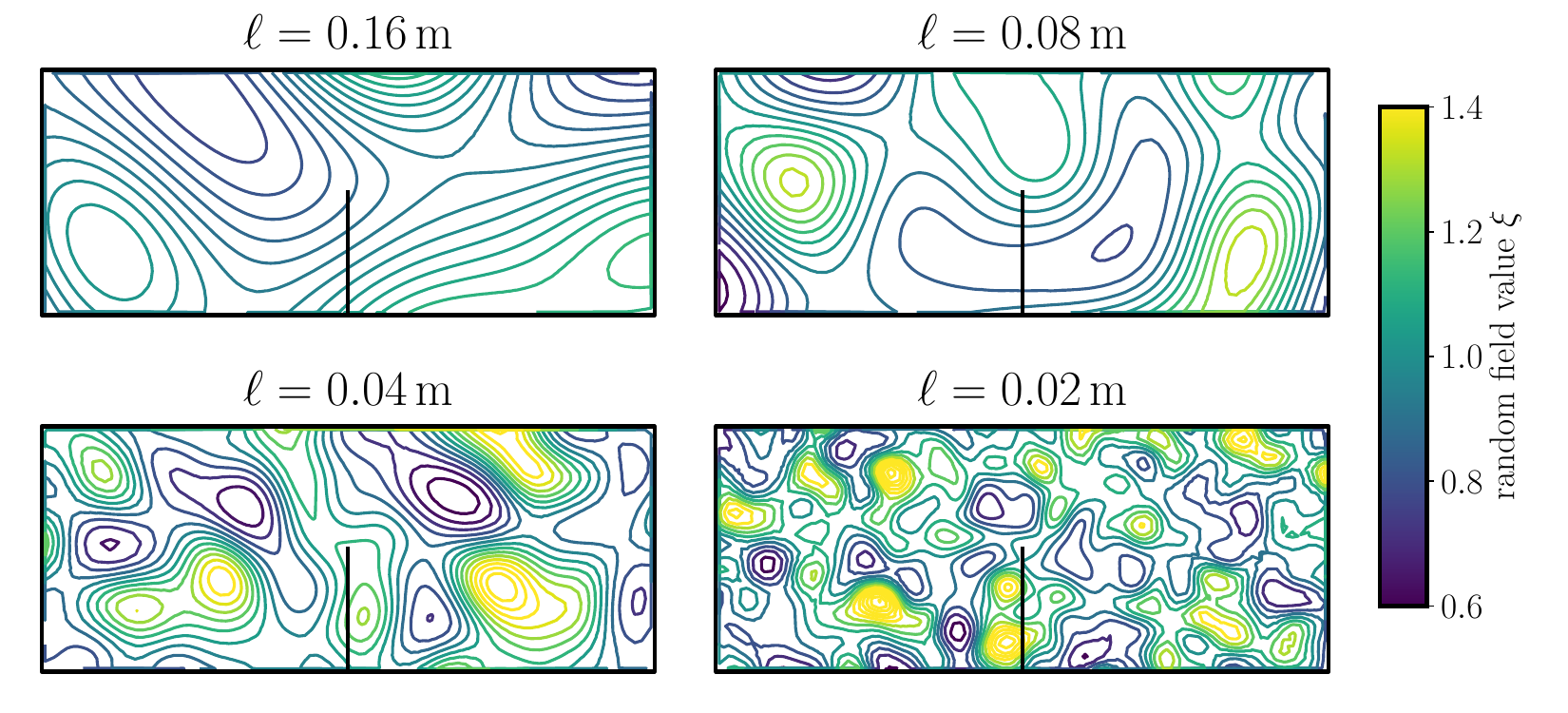}
\caption{Typical realizations of the random field of Young's modulus for different  correlation lengths.} \label{fig:RF}
\end{figure}

Once the equivalent Gaussian random variables are generated, the target lognormal field $\xi(\xx)$ is calculated through Eq.~\eqref{e16}. Fig.~\ref{fig:RF} shows the realizations of the random field $\xi(\xx)$ for different values of correlation length $\ell$. In the present  analysis,  1000 realizations of random field $\xi(\xx)$ are generated for each correlation length. For each realization, we perform the stress analysis, and calculate the energy release rate by Eq.~\eqref{e12} as well as the values of $J_*$ and $J-$integral. From the results of all 1000 realizations, we obtain the mean and standard deviation of these quantities.

\section{Result and Discussion}

 \begin{table}[!b]
\caption{Mean values and CoVs of the energy release rate, $\mathcal{G}$ for different correlation lengths.} \label{tab:G}
\centering
\begin{tabular}{ccc}
\hline
$\ell$ & Mean value & CoV \\ 
{[m]} & [J/m$^2$] &  [-]\\ \hline
0.16 & 86.56 & 0.186 \\
0.08 & 85.90 & 0.165 \\
0.04 & 86.11 & 0.137 \\
0.02 & 85.91 & 0.118 \\\hline 
\end{tabular}
\end{table}

Table~\ref{tab:G} shows mean values and coefficients of variation (CoVs) of the energy release rate calculated from Eq.~\eqref{e12}.  It is seen that the mean energy release rate depends mildly on the correlation length. The mean value obtained from the stochastic analysis is slightly larger than the energy release rate calculated from the deterministic analysis ($\mathcal{G} =$ 83.57\,J/m$^2$). This difference can be explained by the Irwin relationship: $\mathcal{G} = K_I^2/E$, which holds true for each stochastic realization. Note that, since the Young modulus varies spatially, $E$ in the Irwin relationship corresponds to the value of Young modulus at the crack tip \cite{Eis87}. By assuming that the mean  stress intensity factor follows approximately the same expression as Eq.~\eqref{e13}, the mean energy release rate is given by $\langle{\mathcal{G}}\rangle \approx K_I^2 \langle1/E\rangle$. The local Young modulus follows a~lognormal distribution with a~mean value of 30\,GPa and CoV of 0.2. It can be shown that the mean value of $1/E$ at any local point is equal to 0.035\,GPa$^{-1}$, and therefore we have $\langle{\mathcal{G}}\rangle  =$ 86.91\,J/m$^2$. Table ~\ref{tab:G} shows that, as the correlation length increases, $\langle{\mathcal{G}}\rangle$ approaches the result of the foregoing hand calculation. This is because, as the correlation length increases, the spatial variation of Young modulus diminishes for each realization. Therefore, the effect of spatial randomness of Young modulus on the stress intensity factor is minimal. The present calculation is able to predict the exact mean energy release rate. The stochastic analysis shows that, even for small values of correlation length, the mean energy release rate is pretty close to the result predicted by the Irwin relationship (within an~error less than 2\%). This indicates that the mean stress intensity factor is minimally affected by the spatial randomness of Young modulus.  As a~comparison, the deterministic analysis uses the average Young modulus $E = 30$\,GPa, and it yields $\mathcal{G} =$ 83.57\,J/m$^2$. Therefore, we conclude that the difference in the mean behavior of the energy release rate predicted by the stochastic and deterministic analyses can be primarily attributed to the fact that $1/\langle E (x_1, x_2) \rangle \neq \langle 1/E(x_1, x_2) \rangle$.

Table~\ref{tab:G} shows that the CoV of the energy release rate decreases considerably with a~decreasing correlation length. Based on Eq.~\eqref{e12}, the energy release rate can be calculated by ${\mathcal{G}} =\frac{1}{2}\hat{\sig}^2h^2 \pa S(a) /\pa a$, where $S(a) =$ compliance function of the specimen. In the  case where the correlation length is large compared to the specimen size, the spatial variation of Young modulus is minimal and the statistics of $\pa S(a) /\pa a$ is solely governed by the statistics of the inverse of the Young modulus at any local point. When the correlation length decreases, the Young modulus exhibit a~strong spatial variation. The compliance of the specimen represents a~net effect of random compliance at local points. Therefore, the CoV of specimen compliance would decrease when the local compliance at different points becomes more statistically uncorrelated inferred by a~decreasing correlation length.

\begin{table}[!tb]
\caption{Mean values and CoVs of $J-$integral for different contours and correlation lengths.} \label{tab:meanstd}
\centering
\begin{tabular}{rrrrrrrrrr}
\hline
contour & 1&2&3&4&5&6&7&8&9\\ \hline
$\ell=0.16$\,m: mean & 86.52 & 86.52 & 86.52 & 86.51 & 86.52 & 86.52 & 86.53 & 86.54 & 86.54\\
CoV&0.185 & 0.184 & 0.183 & 0.182 & 0.181 & 0.181 & 0.181 & 0.181 & 0.181\\\hline
$\ell=0.08$\,m: mean & 85.85 & 85.84 & 85.83 & 85.84 & 85.86 & 85.88 & 85.91 & 85.93 & 85.96\\
CoV&0.161 & 0.157 & 0.155 & 0.153 & 0.152 & 0.151 & 0.151 & 0.151 & 0.151\\\hline
$\ell=0.04$\,m: mean & 86.01 & 85.98 & 85.99 & 86.02 & 86.08 & 86.14 & 86.20 & 86.25 & 86.30\\
CoV&0.127 & 0.118 & 0.113 & 0.110 & 0.110 & 0.110 & 0.110 & 0.111 & 0.111\\\hline
$\ell=0.02$\,m: mean & 85.73 & 85.68 & 85.70 & 85.76 & 85.85 & 85.90 & 85.92 & 85.93 & 85.94\\
CoV&0.095 & 0.079 & 0.073 & 0.073 & 0.074 & 0.075 & 0.076 & 0.077 & 0.077\\\hline
\end{tabular}
\end{table}

Table ~\ref{tab:meanstd} presents the mean values and the CoVs of $J-$integral computed for 9 different contours (Fig. ~\ref{fig:geometry}). It is seen that, for all values of the correlation length considered here,  the mean value of $J-$integral is independent of the choice of contour, and it is very close to the mean energy release rate calculated from Eq.~\eqref{e12} (Table 1). This result supports our previous conclusion that the mean value of $J-$integral is path independent and it yields the mean energy release rate, i.e.  Eq.~\eqref{e11}.

It is seen from Table~\ref{tab:meanstd} that, for a~given contour, the CoV of $J-$integral decreases with a~decreasing correlation length. The $J-$integral can be rewritten by $J = \int_\Ga P_{1j} n_j\dd s$ where $P_{ij} = \bar{U} \del_{ij} - \sig_{jk} u_{k, i}$. In stochastic analysis, we can express $P_{1j} n_j$ by $\langle P_{1j} n_j \rangle [1+ \zeta(s)]$, where $\zeta(s) =$ a~zero-mean random field defined along the contour. By using this decomposition, the $J$-integral can be written by 
\beq\label{e17}
J = \langle J \rangle + \int_\Ga \langle P_{1j} n_j \rangle  \zeta(s) \dd s
\eeq
where $\langle J \rangle = \int_\Ga \langle P_{1j} n_j \rangle \dd s =$ mean value of $J-$integral. Therefore, the standard deviation of $J$ is governed by the randomness of the second term of Eq.~\eqref{e17}, which is essentially a~weighted sum of $\zeta(s)$ along the contour $\Ga$. When the correlation length of the Young modulus field  is small, the statistics of $\zeta$ at each location along the contour becomes uncorrelated. Consequently the integral $\int_\Ga \langle P_{1j} n_j \rangle  \zeta(s) \dd s$ would exhibit a~smaller standard deviation, which leads to a~smaller CoV of the $J-$integral. On the other hand, Table~\ref{tab:meanstd} shows that, for a~given correlation length, the CoV of $J-$integral decreases and then approaches almost a~constant as the contour gets larger. However, this constant differs from the CoV of the energy release rate. 

By comparing Tables~\ref{tab:G} and \ref{tab:meanstd}, we observe that the $J-$integral could give a~reasonable estimation of the CoV of the energy release rate if the size of domain enclosed by the contour is smaller or at least comparable to the correlation length of the Young modulus. If the domain size is larger than the correlation length,  the CoV of $J$ would differ considerably from the CoV of the energy release rate. As indicated by Tables~\ref{tab:G} and \ref{tab:meanstd}, the difference could be as large as 35\%.

\begin{table}[!tb]
\caption{Mean values and CoVs of $J_*$ for different contours and correlation lengths.} \label{tab:meanstdJ*}
\centering
\begin{tabular}{rrrrrrrrrr}
\hline
contour & 1&2&3&4&5&6&7&8&9\\ \hline
$\ell=0.16$\,m: mean & 86.54 & 86.55 & 86.55 & 86.55 & 86.55 & 86.54 & 86.54 & 86.54 & 86.54\\
CoV&0.187 & 0.187 & 0.187 & 0.187 & 0.187 & 0.187 & 0.187 & 0.187 & 0.187\\\hline
$\ell=0.08$\,m: mean & 85.88 & 85.89 & 85.88 & 85.88 & 85.88 & 85.88 & 85.88 & 85.88 & 85.88\\
CoV&0.166 & 0.166 & 0.166 & 0.165 & 0.165 & 0.165 & 0.165 & 0.165 & 0.165\\\hline
$\ell=0.04$\,m: mean & 86.10 & 86.10 & 86.09 & 86.08 & 86.09 & 86.09 & 86.09 & 86.09 & 86.09\\
CoV&0.140 & 0.140 & 0.139 & 0.139 & 0.139 & 0.138 & 0.138 & 0.138 & 0.138\\\hline
$\ell=0.02$\,m: mean & 85.92 & 85.91 & 85.90 & 85.89 & 85.90 & 85.90 & 85.90 & 85.90 & 85.90\\
CoV&0.124 & 0.122 & 0.121 & 0.120 & 0.120 & 0.120 & 0.120 & 0.120 & 0.120\\\hline
\end{tabular}
\end{table}

Table~\ref{tab:meanstdJ*} shows the mean and CoV of $J_*$. It is seen that, for all different values of the correlation length, the mean value of $J_*$ is path independent, and is essentially equal to the mean energy release rate calculated from the complementary potential energy function (Table ~\ref{tab:G}). Meanwhile, the CoV of $J_*$ is almost independent of the choice of the contour. The small difference in CoV of $J_*$ for different contours (less than 3\%) arises from the numerical approximation of the domain integration. It is seen that the CoV of $J_*$ is very close to the CoV of the energy release rate calculated from the complementary potential energy function. This result supports the theoretical derivation that $J_*$ indeed yields the energy release rate of a~non-homogeneous elastic medium.

\section{Conclusion}
This study investigates the application of path integral for calculating the energy release rate of  an~elastic medium with spatially random Young modulus. The analysis shows that the mean value of $J-$integral is path independent and it yields the exact mean value of the energy release rate when the random field of the elastic modulus is homogenous. It is also demonstrated that the mean energy release rate predicted by the stochastic analysis is slightly different from that predicted by the deterministic analysis using the mean value of the elastic modulus. 

The CoV of the $J-$integral is generally path dependent. It is shown that the CoV of $J-$integral converges to the CoV of the energy release rate when the domain enclosed by the contour is smaller than the correlation length of the random field of the elastic modulus. Otherwise, the $J-$integral would grossly underestimate the CoV of the energy release rate. In order to predict the statistics of the energy release rate with an~arbitrary  choice of contour, a~domain integral needs to be added to the $J-$integral. 

The present analysis also shows that the correlation length of the random field of the elastic modulus has a~minimal effect on the mean value of the energy release rate. However, it has a~strong effect on the second statistical moment of the energy release rate. For a~given specimen, the CoV of the energy release rate decreases considerably as the correlation length decreases.

\section*{Acknowledgment} J.-L. Le gratefully acknowledges funding support from the U.S. National Science Foundation (Grant CMMI-2151209) to the University of Minnesota. Authors from the Brno University of Technology gratefully acknowledges financial support from the Czech Science Foundation (Grant GA24-11845S) and the internal Brno University of Technology grant (No.~FAST-J-23-8323) supporting students.

\section*{Data Availability} The individual values of energy release rate $\mathcal{G}$, $J$-integral and $J_*$-integral for each realization and contour are available at DOI \href{https://doi.org/10.5281/zenodo.14579387}{\texttt{10.5281/zenodo.14579387}}.

\printbibliography

\end{document}